# Fundamental measure theory of hydrated hydrocarbons

VICTOR F. SOKOLOV AND GENNADY N. CHUEV

*Institute of Theoretical and Experimental Biophysics, Russian Academy of Sciences, Pushchino, Moscow region, 142290 Russia*

+7-0967-739-109
+7-0967-790-553

vicvor@mail.ru

To calculate the solvation of hydrophobic solutes we have developed the method based on the fundamental measure treatment of the density functional theory. This method allows us to carry out calculations of density profiles and the solvation energy for various hydrophobic molecules with a high accuracy. We have applied the method to the hydration of various hydrocarbons (linear, branched and cyclic). The calculations of the entropic and the enthalpic parts are also carried out. We have examined a question about temperature dependence of the entropy convergence. Finally, we have calculated the mean force potential between two large hydrophobic nanoparticles immersed in water.

*Hydrophobic interactions, solvation energy, complex formation*

SPT – Scaled Particle Theory

WPI – Widom particle insertion

DFT – Density Functional Theory

FMT – Fundamental Measure Theory

LJ – Lennard-Jones

HS – Hard sphere

# Introduction

Membrane formation, protein folding, stabilization of various biomacromolecular complexes including nucleic acids and lipids are largely driven by hydrophobic interactions [1, 2]. The main reason of it is that all these substances contain a large number of nonpolar groups. Hydrophobic hydration describes structural and dynamical changes of water around a single nonpolar solute whereas hydrophobic interaction forms the net of intermolecular bonds (e.g. hydrogen bonds for water). This latter effect refers to a tendency of nonpolar solutes to stick together in aqueous solutions. The theory of the hydrophobic effect is intrinsically difficult. Despite of long history of research of hydrophobic interactions, at present there is no complete understanding of nature of hydrophobic interactions. Difficult multiscale character of these effects can be revealed as well as microscopic changes of water structure near small hydrophobic groups and conformations or aggregation of biomacromolecules at mesoscopic scales up to several tens of angstrom [3].

Historically, hydrophobicity has been described as low solubility of nonpolar solutes in water in comparison with organic solvents [4]. During long time hydrophobic hydration has been regarded as a peculiar type of solvation, since it is characterized by an unfavorable entropy contribution rather than by an unfavorable enthalpy change [4, 5], and the iceberg-like models postulating of the increasing water structure around of the solute have been invoked to explain this property. It has been specified subsequently that polar solutes also reveal unfavorable hydration entropies [6, 7].

A wide variety of approaches has been used to describe and explain hydrophobic hydration on the molecular level. Earlier approaches such as scaled particle theory (SPT) [8-10] and the methods of the integral equations [11, 12] were essentially analytical. Later, an increase of the computation capability has been made possible computer simulations of solvation process [13]. The coupling of such simulations with statistical thermodynamic techniques, such as the free energy perturbation method [14] and the Widom particle insertion (WPI) approach [15], provide computation of free energy of hydration.

The immersion of a solute (e.g. any hydrocarbon) to water strongly modifies its local structure. This effect is very significant and application the computer simulation for its description is often necessary. Molecular dynamics and Monte Carlo techniques are most frequently used for modeling molecular interactions in solutions [16]. However, their applications to the problem of hydrophobic interactions of macromolecules demand huge computing expenses and in some cases essentially limited because of the multiscale character of these interactions. In last decades the new methods based on a statistical treatment are in progress [3, 17-20]. In our opinion the most suitable method for research of the specified effects is the density functional theory (DFT) [21, 22]. This approach considers liquids as spatially disorder ensembles and also allows one to calculate solvation effects, using the data about the molecular structure and intermolecular interaction potentials. The basic purpose of approach is to construct the free energy functional of the system, which depends on density distribution of liquid particles and interaction potentials between solute and solvent molecules. Within the framework of this approach there are a lot of various models connected with a concrete choice of density functional [23-26]. One of most suitable versions DFT for calculation of solvation is the fundamental measure theory (FMT) [24, 25, 27, and 28]. The FMT determines the free energy functional as the sum of the weighted contributions dependent on geometrical characteristics of a solvated particle. It automatically results in the definition of weighted functions responsible for the volume and the surface contributions to solvation energy.

In the given work we will calculate the hydrophobic solvation on the basis of the FMT. In particular, we will evaluate the size dependence of the excess chemical potential of solute and calculate the excess chemical potentials for hydrocarbons in water. On the FMT basis we will decompose the excess chemical potential into the excess solvation enthalpy and the excess solvation entropy. Moreover, we will estimate the energy of association and the mean force potential between the two solutes within the limit of low concentration of the dissolved particles.

## Fundamental measure theory

The density functional theory (DFT) [21] solves a problem by a search of the free energy for liquids. The free energy consists of two contributions: ideal ($F_{id}[\rho]$) and excess ($F_{ex}[\rho]$) free energies:

$$F[\rho] = F_{id}[\rho] + F_{ex}[\rho], \quad \beta F_{id}[\rho] = \int \rho(\mathbf{r}) \ln[\rho(\mathbf{r})\Lambda^3 - 1] d\mathbf{r}, \qquad (1)$$

where $\Lambda$ is the De-Broglie wave length, $\rho(\mathbf{r})$ is local density, $\beta = (k_B T)^{-1}$ is the inverse temperature, and $k_B$ is the Boltzmann constant.
Minimization of the functional leads to the equilibrium density:

$$\rho(\mathbf{r}) = \rho_b \exp[-\beta U_{ext}(r) + \delta F_{ex}[\rho]/\delta\rho(\mathbf{r}) - \delta F_{ex}[\rho_b]/\delta\rho(\mathbf{r})], \qquad (2)$$

where $U_{ext}(r)$ is intermolecular potential, $\rho_b = \rho(\mathbf{r} \to \infty)$ is bulk density. Thus, if the functional $F_{ex}[\rho]$ is known we can calculate the density profile $\rho(\mathbf{r})$ and then all required characteristics of solvation.

There are many ways of designing the functional $F_{ex}[\rho]$, most of them use the data on functional derivatives $\partial F_{ex}/\partial\rho(\mathbf{r})$ and $\partial^2 F_{ex}/\partial\rho(\mathbf{r})\partial\rho(\mathbf{r}')$ of the particle density. One of the methods is the FMT [24, 25, 27, 28] based on the following representation. It assumes the calculation of the free energy for a non-uniform liquid $F_{ex}[\rho]$, using the information about a homogeneous liquid $F[\rho_b]$. It is possible to do it, since the local density of the non-uniform liquid is practically constant in microvolume whose size does not surpass the characteristic size of liquid particles. Since the local density is different in various microvolumes the free energy can be expressed through integral depending on the density of the free energy of a non-uniform liquid:

$$\beta F_{ex}[\rho] = \int \Phi[n_i(\mathbf{r})] d\mathbf{r}, \qquad (3)$$

where variables $n_i(\mathbf{r})$ are determined as weights

$$n_i(\mathbf{r}) = \int \rho(\mathbf{r}') w_i(\mathbf{r} - \mathbf{r}') d\mathbf{r}' \qquad (4)$$

of the density $\rho(\mathbf{r}')$ averaged with weight factors $w_i(\mathbf{r} - \mathbf{r}')$. Here, index $i$ denotes the number of weight density. These weights are the characteristic functions, determining geometry of the solute, and in the three-dimensional case they are expressed as:

$$w_3(r) = \Theta(\sigma/2 - r), \; w_2(r) = \delta(\sigma/2 - r), \; w_1(r) = w_2(r)/(2\pi\sigma), \qquad (5)$$

$$w_0(r) = w_2(r)/(\pi\sigma^2), \; \mathbf{w}_{v2}(\mathbf{r}) = \mathbf{r}\delta(\sigma/2 - r)/r, \; \mathbf{w}_{v1}(\mathbf{r}) = \mathbf{w}_{v2}(\mathbf{r})/(2\pi\sigma),$$

where $\delta(r)$ and $\Theta(r)$ are the Dirac delta-function and the Heaviside function respectively, $\sigma$ is the diameter of a solvent particle. The weight factors $w_0$, $w_1$, $w_2$, $w_3$ are scalars, $\mathbf{w}_{v1}$ and $\mathbf{w}_{v2}$ are vectors.

Within the framework [24, 25, 27, 28], the function $\Phi[n_i]$ is determined through weight densities $n_0(\mathbf{r}) - n_3(\mathbf{r})$, $\mathbf{n}_{v1}(\mathbf{r})$ and $\mathbf{n}_{v2}(\mathbf{r})$ as

$$\Phi[n_i] = -n_0 \ln(1 - n_3) + (n_1 n_2 - \mathbf{n}_{v1}\mathbf{n}_{v2})/(1 - n_3) + (n_2^3 - 3n_2 \mathbf{n}_{v2}\mathbf{n}_{v2})/(24\pi(1 - n_3)^2). \qquad (6)$$

Using equations (2) and (6), we obtain the equilibrium density

$$\rho(\mathbf{r}) = \rho_b \exp\{-\beta U_{ext}(r) + \sum_i \int \{[\delta\Phi(\mathbf{r}) - \delta\Phi(\mathbf{r}' \to \infty)]/\delta n_i(\mathbf{r})\} w_i(\mathbf{r} - \mathbf{r}') d\mathbf{r}'\}. \qquad (7)$$

In turn the excess part of thermodynamic potential $\Delta\mu_{ex}$ determining the solvation free energy is calculated as

$$\Delta\mu_{ex} = F_{ex} - \sum_i n_i \frac{\delta F_{ex}}{\delta n_i}. \qquad (8)$$

It is significant that integration of the equation (3) for calculation the excess chemical potential (8) have to be carried out from chosen $r_{cut}$ to infinity, but not from 0 to infinity. The main reason of it is the fact that FMT theory in origin does not treat pressure correctly. As a result, the excess chemical potential is proportional to volume of solute and surface part of the excess chemical potential becomes much less than volume part. In this case we can not calculate the excess chemical potential correctly. We have found a way to solve the problem. We have chosen the point $r_{cut}$ so as to predict the experimental surface tension of water 102 cal/(mol Å$^2$). In this case the magnitude of pressure ($\beta p \sigma^3$) is about 0.05. We note that originally the FMT method was advanced for a liquid of hard spheres. The FMT method can be applied to spheroids of rotation [29], and also for various repulsive and attractive potentials [30-32]. In last case the attraction part of potential $U_{att}(r)$ is considered as a perturbation, which gives the contribution to free energy

$$F_{vv}[\rho] = \frac{1}{2} \iint [\rho(\mathbf{r}) - \rho_b] U_{att}(r) [\rho(\mathbf{r}') - \rho_b] d\mathbf{r} d\mathbf{r}'. \qquad (9)$$

When we considered the hard sphere solute which dissolved in Lennard-Jones (LJ) water, we had to take into account the solute-solvent attractive interactions. The first order correction of the excess chemical potential is expressed as

$$F_{uv}[\rho] = \rho_b \int g_{HS}(\mathbf{r}) U_{att}(r) d\mathbf{r}. \qquad (10)$$

Since experiments are most commonly done at fixed pressure $p$, it is convenient to decompose the hydration chemical potential into the excess solvation entropy $\Delta S$ and the excess solvation enthalpy $\Delta H$, achieved by the use of an isobaric temperature derivative [33-36]

$$\Delta S = \left( \frac{\partial \Delta \mu_{ex}}{\partial T} \right)_p, \qquad (11)$$

$$\Delta H = \Delta \mu_{ex} + T \Delta S.$$

The information on weighted densities also allows us to calculate the mean force potential $A_{ass}(\mathbf{r})$ determining the force of interaction between two solutes in an indefinitely diluted solution

$$A_{ass}(\mathbf{r}) = V_{uu}(\mathbf{r}) - \int \sum_i \{[\delta \Phi(\mathbf{r}') - \delta \Phi(\mathbf{r}' \to \infty)]/\delta n_i(\mathbf{r}')\} w_i(\mathbf{r} - \mathbf{r}') d\mathbf{r}', \qquad (12)$$

where $V_{uu}(\mathbf{r})$ is the direct intermolecular interaction potential.

## Results and Discussions

Using the fundamental measure theory, we have calculated the radial distribution functions for linear, branched and cyclic hydrocarbons (methane, ethane, butane and et al.) in water [see 38]. On the basis of them and the equation (8) we have evaluated the excess chemical potential for the hydrocarbons in water. The repulsive interaction between solute and solvent molecules with $\sigma = \sigma_v = 2.77$ Å and $\varepsilon_v = 0.1554$ kcal/mol [39] is simulated as hard spheres (HS). The sizes of hydrocarbons $\sigma_u$ and their Lennard-Jones parameters $\varepsilon_u$ were extracted from [37, 40]. The water density $\rho_b \sigma_v^3$ was equal to 0.7. The calculations of the attractive contribution have been carried out in the framework of a perturbation theory to take into account the most factors which influence on a solvation process. The attractive interactions between solute and solvent molecules have been calculated as the first order corrections to the excess chemical potential (10) for attractive part of Lennard-Jones potential, according Weeks-Chandler-Anderson derivation

$$U_{att}(r) = \begin{cases} -\varepsilon_{uv}, & r < \sqrt[6]{2}\sigma_{uv}, \\ 4\varepsilon_{uv}\left(\left(\dfrac{\sigma_{uv}}{r}\right)^{12} - \left(\dfrac{\sigma_{uv}}{r}\right)^{6}\right), & r \geq \sqrt[6]{2}\sigma_{uv}, \end{cases} \qquad (13)$$

where $\sigma_{uv} = \dfrac{\sigma_u + \sigma_v}{2}$ and $\varepsilon_{uv} = \sqrt{\varepsilon_u \varepsilon_v}$.

In addition, we have calculated the attractive contribution to the energy of solvent-solvent interactions using equation (11). Note that this attractive part of the solvent-solvent energy decreases from 14 to 1 percent with increasing the solute radius from 0.5 to 20 Å. Opposite, the solute-solvent contribution increase with increasing the radius of solute. The magnitude of solute-solvent energy is about 50% or more of the excess chemical potential. It's not necessary to calculate the second order of perturbation theory because of the size of hydrocarbons is enough large and this contribution is negligible.

Figure 1 shows the difference between the FMT and experimental results for the excess chemical potentials, enthalpies and entropies of hydrocarbons. The magnitudes of the excess chemical potentials for small hydrocarbons (methane, ethane and propane) are bit different from experimental results. But the difference of the excess chemical potentials for large solutes is more essential. The main reason of it consists in that the methane, ethane and propane can be easily approximate by hard spheres solutes without loss of accuracy. Opposite for the large solute, it is necessary to take into account its molecular structures, although the FMT results are enough close to the experimental results.

Moreover, we carried out the analysis of decomposition the excess chemical potential for hydrocarbons on entropic and enthalpic parts. Using Eqn. (14), we have decomposed the excess chemical potential to two parts $\Delta H$ and $-T\Delta S$. The calculated and the experimental excess chemical potential, the enthalpies, and the entropies of the hydrocarbons from methane to hexane, of the branched hydrocarbons isobutane, 2-methylbutane, and neopentane, and of the cyclic hydrocarbons cyclopentane and cyclohexane are shown in Figure 1. The obvious feature is that the entropic and enthalpic terms are larger in absolute value than the excess chemical potential. The hydration enthalpies are large and favorable, and the hydration entropies are large and unfavorable. The entropic terms are marginally larger in absolute value than the enthalpic terms resulting in the unfavorable but small hydration free energies of the hydrocarbons. It is recognized that this behavior is typical for hydrophobic hydration. Solvation of apolar compounds in most other solvents, in fact, is usually accompanied by smaller enthalpic and entropic changes. Figure 1 shows that difference between the experimental and the calculated enthalpy and entropy parts of the excess chemical potential [41]. The discrepancies in enthalpy do not exceed 7 kcal/mol for all hydrocarbons (except the 2-methylbutane and cyclopentane due to strong difference the molecules from spheres). The calculated excess chemical potential of methane is 3% more positive than the experimental value ($2.01$ kcal/mol). Partially, this difference is due to the fact that the calculated hydration enthalpy of methane is about 3.97 kcal/mol more negative than the experiments (-2.75 kcal/mol). This represents a significant discrepancy if compared to the better agreement between the calculated and the experimental relative hydration enthalpies for the higher alkanes. The calculated hydration entropy of methane, although in better agreement with the experimental hydration entropy ($-T\Delta S = 6.04$ kcal/mol), is found to be too large. The difference in the enthalpy and entropy partially cancel each other resulting in a smaller discrepancy in the excess chemical potential. The calculated and experimental hydration free energies of the alkanes are positive and practically do not increase with solute size (Figure 1).

The decreasing in the excess chemical potential going from methane to ethane is well reproduced by the calculations. The calculations overestimate the value of negative enthalpy change and too highly underestimate entropy loss in going from methane up to ethane. Both effects bring in less favorable hydration free energy of ethane. It appears, because current hydrocarbons model should take into account a larger benefit in favorable hydrocarbon-water interactions in going from methane up to ethane without the further loss of entropy.

The experiments show that each methylation beyond the first increases the free energy of hydration of the linear alkanes by approximately 0.2 kcal/mol whereas the calculated free energy increase at each methylation is on average a little bit higher. Unfortunately, the calculations do not reproduce the experimental 1.2 kcal/mol free energy difference between hexane and cyclohexane and 1.1 kcal/mol between cyclopentane and pentane.

The calculated and experimental enthalpies of hydration of the normal alkanes are not in good agreement (Figure 1). The calculations overestimate the magnitude of the experimental enthalpies of hydration. Despite the quite favorable enthalpies of hydration, the solubility in water of these compounds remains low due to the unfavorable entropies. Unfortunately, the calculated and the experimental enthalpies of hydration of the linear alkanes are not in good agreement too. The experimental entropy losses are underestimated by the calculations. This effect could be in partly due to having ignored the torsional degrees of freedom of the carbon chains. For linear alkanes each additional methylation decreases $T\Delta S$ by about 2.0 kcal/mol. The $T\Delta S$ values are larger in magnitude than the $\Delta H$ values resulting in the positive hydration free energies discussed above.

The free energies of hydration of the cyclic alkanes favor hydration less than their corresponding linear homologues. The free energy drop caused by cyclization is reproduced by the calculations, although a little overestimated. It has been shown experimentally and computationally that the solubility of cyclohexane increased with respect to hexane significantly more favorable than expected enthalpy of hydration. Despite their different accessible surface areas, the enthalpies of cyclohexane and hexane are of similar magnitude. The calculations also show that the larger solubility of cyclopentane compared to an alkane of the same solvent accessible surface area is due mostly to a more favorable hydration enthalpy, although partially offset by a more unfavorable entropic component.

The hydrophobic effect is frequently connected to characteristic temperature dependences [42-44]. One of the most surprising observations is the entropy of transition convergence of nonpolar molecules from gas phase or nonpolar solvent into water at temperature about 400 K to approximately zero entropy change. We have made the calculations to indicate that the FMT is able to predict the temperature convergence of entropy as well as qualitatively and quantitatively correctly. Computer simulations were carried out to calculate the water-oxygen radial distribution function $g(r)$ and the excess chemical potential for hydrocarbons at several temperatures along the experimental saturation curve of water.

Using equation (11) we have obtained the solvation entropy by calculating the derivative of the chemical potential along the saturation curve. Figure 2 shows the temperature dependence of the entropy $\Delta S$ for the different solutes for two cases of calculation. In the first case (Fig.2a) we calculated the excess chemical potential without the fact that the diameter of water decreases with temperature increasing. The entropies are large and negative at room temperature for all the solutes and decrease in magnitude with increasing temperature. The temperature dependence of entropies is approximately linear with slopes increasing with the increasing solute size. Moreover, the entropies converge at about 400 K to approximately zero entropy, although at closer inspection the temperature range of the convergence region is several 10 K and the entropy is not exactly zero at convergence. In the second case, we have taken into account the dependence of the solvent diameter on temperature [45]. Figure 2b shows that the point of entropy convergence has shifted to region where temperature and entropy magnitude is about 470 K and -2.5 cal/(mol K), correspondingly. The convergence region has become a bit wider. It is significant that taking into account the contributions solute-solvent interactions has changed both the point of entropy convergence (about 500 K) and width of convergence region.

We have to note the one more benefit of fundamental measure theory. The theory allows calculating the mean force potential between two solutes surrounded with solvent particles. In the work we have obtained the profiles of mean force potentials of different solutes at various temperatures. The example of the curve is plotted on figure 3. In the upper-right corner of the figure 3 is plotted the localization dependence maxima and minima on temperature. Figure 3 shows that the extremums of mean force potential linearly move to left with increasing the temperature. In reality the localization of extremums does not practically change with temperature increasing. The reason of discrepancy between calculated and simulation results consist in a small error in calculations by FMT. The magnitude of potential barrier obtained by FMT a little underestimate the simulation result (about 0.1 kcal/mol).

## Conclusions

In this work we used the fundamental measure theory to the quantitative description of the hydrophobic phenomena based on the density functional theory. As a result we have received profiles of radial distribution functions for the isolated solutes (hydrocarbons) in hard sphere fluid. Using the profile of a distribution function, in frameworks of FMT we have calculated the excess chemical potentials for linear, branched and cyclic hydrocarbons in water by modeling of them as spherical solutes. The water molecules have been modeled as LJ solvent. In addition, we have derived the excess chemical potentials for hydrocarbons to entropic and enthalpic part. The differences between the experimental and the obtained from FMT results are not considerable for hydration free energy and more essential for entropy and enthalpy. Moreover, we have calculated the mean force potential for methane-like solute and have shown the changing the extremums with temperature increasing.

**Acknowledgements.** This work was partly supported by the Russian Foundation of Basic Research.

Fig.1. The calculated and the experimental excess chemical potentials (solid lines), enthalpies (dash-dotted lines), and entropies (dashed lines) of hydrated alkanes: (circles) normal alkanes from $C_1$ to $C_6$ in order of increasing solute diameter; (triangles-up) isobutane; (rotated triangle) 2-methylbutane; (triangles-down) neopentane; (squares) cyclopentane; (diamonds) cyclohexane.

Fig.2. Hydration entropy of HS solutes with radii corresponding hydrocarbons (see legend) as a function of temperature along the saturation curve of water. Case 2a corresponds to the water HS diameter independent on temperature, while the case 2b to the water diameter depending on temperature [45].

Fig.3. The mean force potential $A_{ass}(r)$ of a methane-like solute at 300 K ($\sigma_{uv} = 3.336$ Å, $\varepsilon_{uv} = 0.241$ kcal/mol, $\sigma_u = 3.57$ Å, $\varepsilon_u = 0.4$ kcal/mol). The temperature dependencies of the first minimum and maximum are represented in the upper-right corner of the figure.

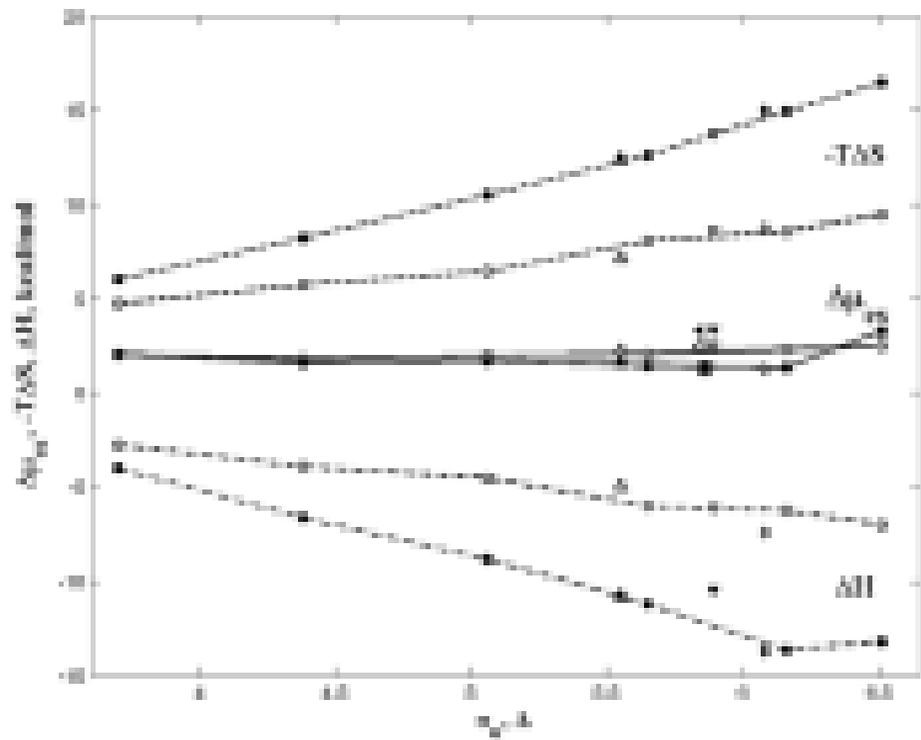

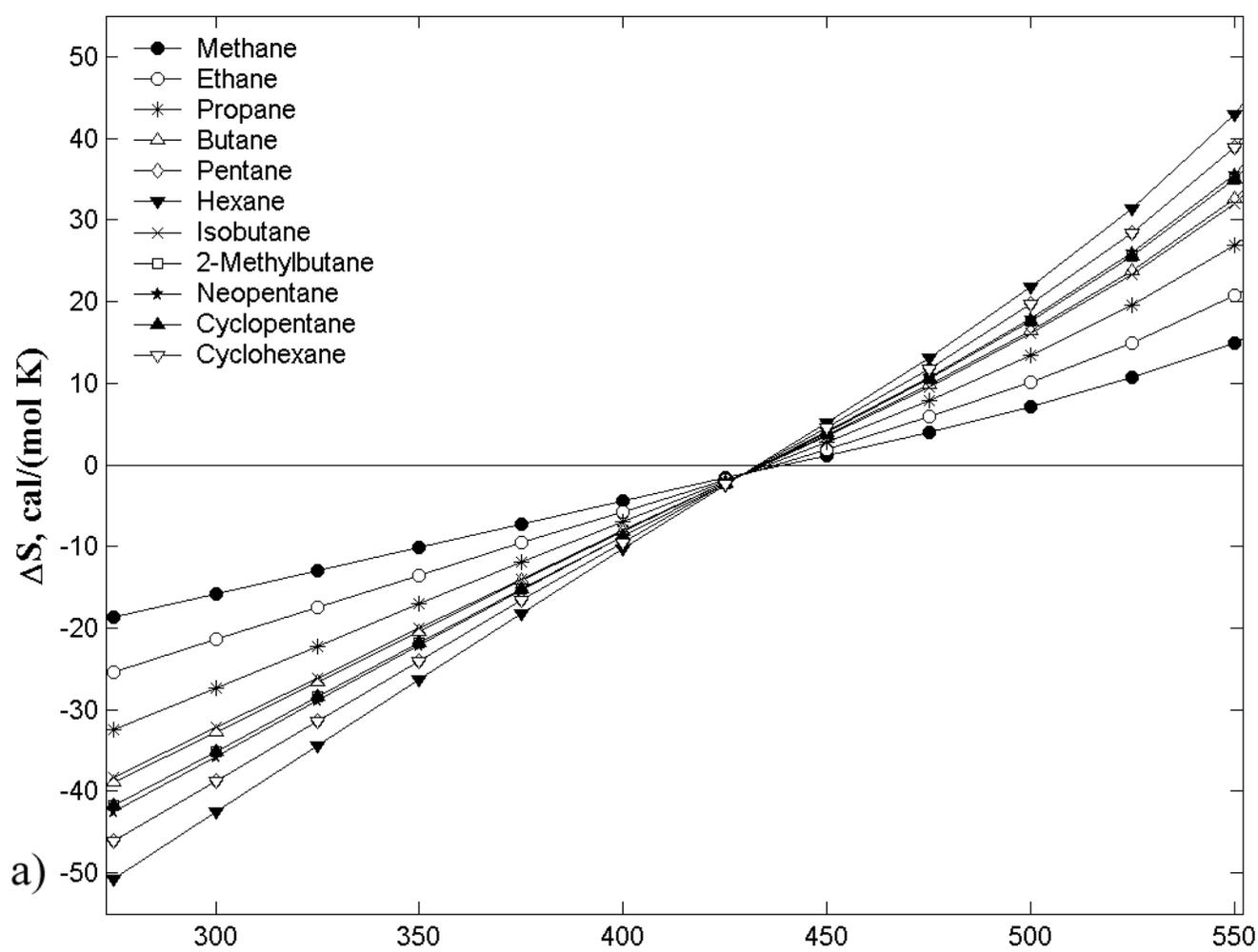

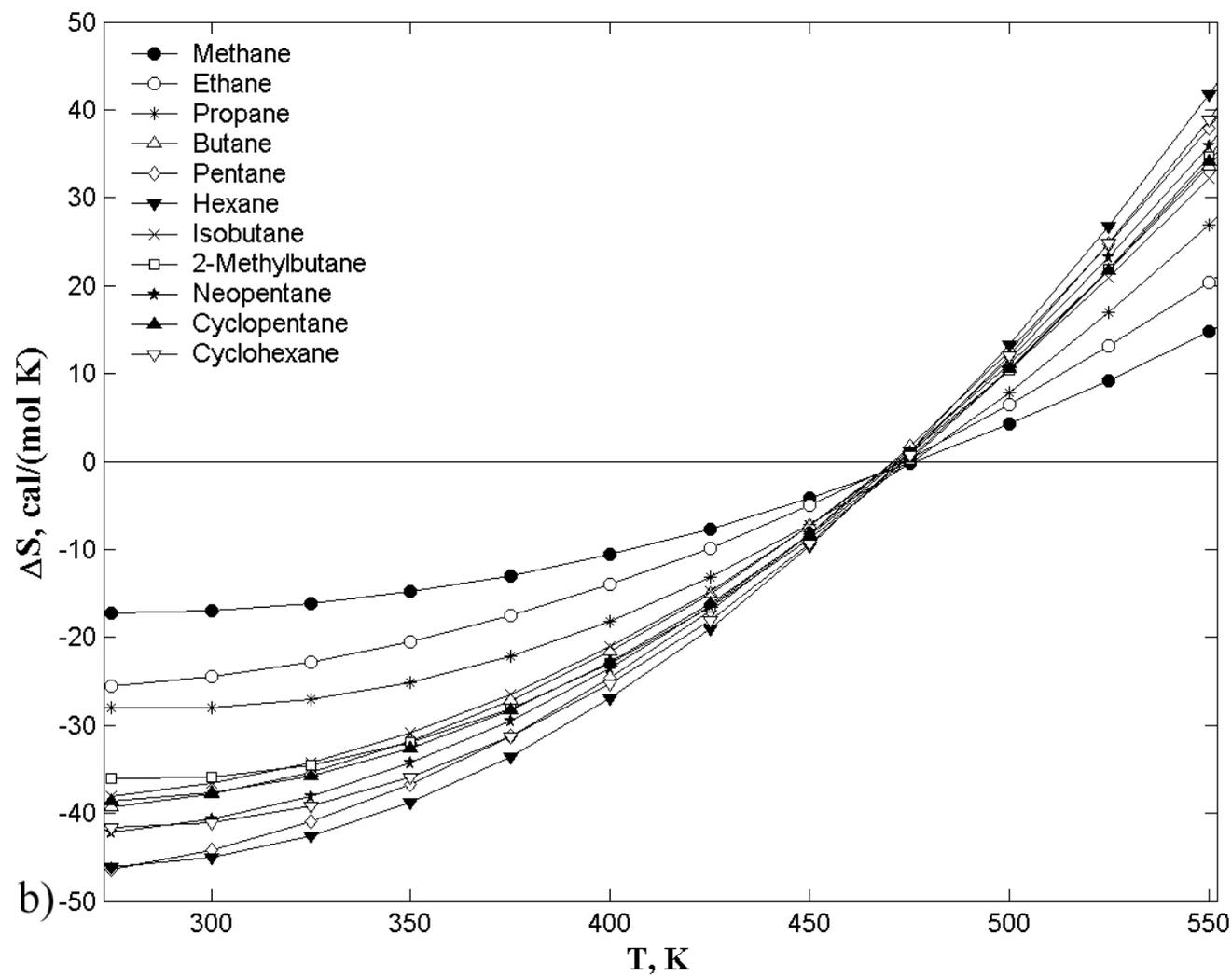

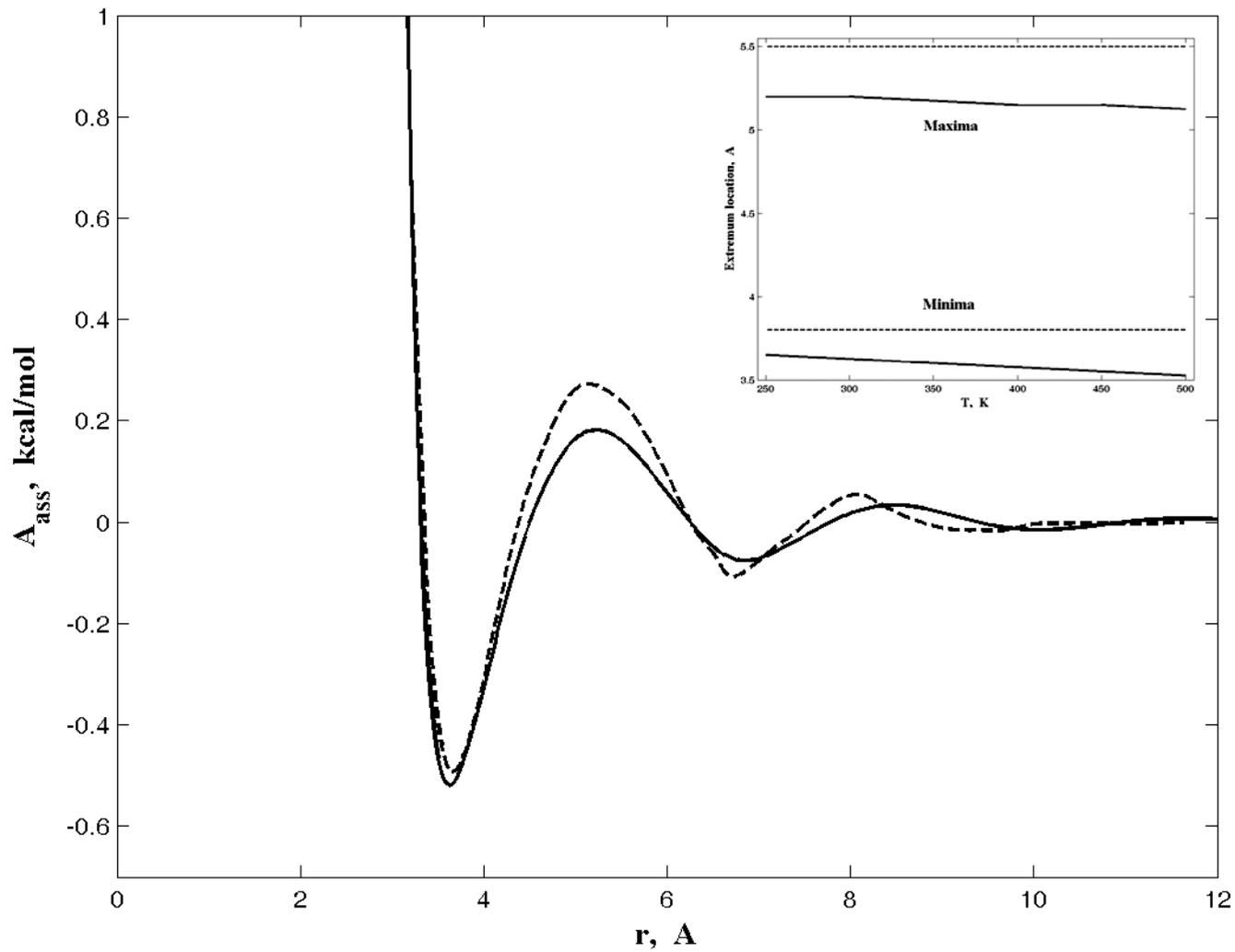